\documentclass[twocolumn,prd,showpacs,floatfix,preprintnumbers,nofootinbib,a4paper]{revtex4}

\usepackage{epsfig}

\def\lsim{\mathrel{\raise.3ex\hbox{$<$\kern-.75em\lower1ex\hbox{$\sim$}}}}
\def\gsim{\mathrel{\raise.3ex\hbox{$>$\kern-.75em\lower1ex\hbox{$\sim$}}}}
\def\sigmsd{\sigma_{SD}}

\def\sigv{\langle\sigma v\rangle}
\def\nuflub8{\phi^\nu_B}
\def\nuflube7{\phi^\nu_{Be}}
\def\epstrans{\epsilon_{trans}}
\def\epsnuc{\epsilon_{nuc}}

\def\lesssim{\buildrel < \over {_{\sim}}}
\def\gtrsim{\buildrel > \over {_{\sim}}}

\def\rchi{r_\chi}
\def\rhochi{\rho_\chi}
\def\mchi{m_\chi}
\def\Tchi{T_\chi}
\def\Tb{T_b}

\def\ntot{n_{tot}}
\def\Ltrans{L_{trans}}
\def\Lnuc{L_{nuc}}
\def\rnuc{r_{nuc}}

\begin{document}

\title{Main sequence stars with asymmetric dark matter}

\author{Fabio Iocco$^{1,2}$, Marco Taoso$^{3^4}$, Florent Leclercq$^{1,5}$ and Georges Meynet$^{6}$}

\affiliation{$^{1}$ Institut d'Astrophysique de Paris, UMR 7095, CNRS, UPMC Univ. Paris 06, 98bis Bd Arago, F-75014 Paris, France} 
\affiliation{$^{2}$ The Oskar Klein Center for Cosmoparticle Physics, Department of Physics, Stockholm University, AlbaNova, SE-106 91 Stockholm, Sweden} 
\affiliation{$^{3}$ Multidark fellow at IFIC (CSIC-Universitat de Valencia), Ed.Instituts, Apt.22085, 46071 Valencia, Spain}
\affiliation{$^{4}$ Department of Physics and Astronomy, University of British Columbia, Vancouver, BC V6T 1Z1, Canada}
\affiliation{$^{5}$ \'Ecole Polytechnique ParisTech, Route de Saclay, F-91128 Palaiseau, France}
\affiliation{$^{6}$ Geneva Observatory, University of Geneva, Maillettes 51, 1290 Sauverny, Switzerland}

%\affiliation{Institut d'Astrophysique de Paris, UMR 7095-CNRS, Universit\'e Pierre et Marie Curie, 98 bis Boulevard Arago 75014, Paris, France}

\date{\today}

%%%%%%%%%%%%%%%%%%%%%%%%%%%%%%%%%%%%%%%%%%%%%%%%%%%%%%%%%%%%%%%%%%%%%%
\begin{abstract}
%%%%%%%%%%%%%%%%%%%%%%%%%%%%%%%%%%%%%%%%%%%%%%%%%%%%%%%%%%
We study the effects of feebly or non-annihilating weakly interacting Dark Matter (DM) particles on stars
that live in DM environments denser than that of our Sun. We find that 
the energy transport mechanism induced by DM particles
can produce unusual conditions in the core
of Main Sequence stars, with effects which can 
potentially be used to probe DM properties.
We find that solar mass stars placed in DM densities 
of $\rhochi$ $\geq$ 10$^2$GeV/cm$^3$
are sensitive to Spin-Dependent scattering
cross-section $\sigmsd$ $\geq$ 10$^{-37}$cm$^2$ and a 
DM particle mass as low as $\mchi$=5 GeV, 
accessing a parameter range weakly constrained 
by current direct detection experiments.
%%%%%%%%%%%%
\end{abstract}
%%%%%%%%%%%%%%%%%%%%%%%%%%%%%%%%%%%%%%%%%%%%%%%%%%%%%%%%%%%%%%%%%%%%%%

\maketitle

%%%%%%%%%%%%%%%%%%%%%%%%%%%%%%%%%%%%%%%%%%%%%%%%%%%%%%%%%%%%%%%%%%%%%%
{\it Introduction---}
%%%%%%%%%%%%%%%%%%%%%%%%%%%%%%%%%%%%%%%%%%%%%%%%%%%%%%%%%%%%%%%%%%%%%%
Identifying the Dark Matter (DM) is one of
the most enthralling problems of modern cosmology and particle
physics. 
Weakly Interacting Massive Particles (WIMPs) are among the 
most popular candidates and are currently searched for with
different strategies, see e.g Ref.~\cite{reviews} for reviews.
The effects of WIMPs on stars have been 
recently investigated as possible additional probes for
DM searches: in presence of self-annihilations,
WIMPs captured inside a star can provide an exotic
source of energy.
Whereas the effects on the Sun are negligible because of the small
DM density in the solar neighborhood,
they may be remarkable on stars living
in dense DM environments such as the Galactic Center,
or the first halos to harbor star formation at high redshift
\cite{Scott:2008ns,Moskalenko:2007ak,Iocco:2008xb,Taoso:2008kw}. 
In general, indirect searches of DM would fail for
very feebly annihilating DM candidates. This is the case, for instance,
of many asymmetric DM  (ADM) models (see e.g.\cite{ADMreviews}), where
DM annihilations may not occur in presence of an asymmetry between DM
particles and antiparticles.
Still, these particles accumulated in a star can scatter off
nuclei and transport energy within its bosom.
This effect results in a modification of
the density and temperature profiles which can lead
to detectable changes of
the solar neutrino fluxes and gravity modes
\cite{Dearborn:1990mm,Frandsen:2010yj,Taoso:2010tg,Cumberbatch:2010hh};
more compact objects like neutron stars
may cumulate enough particles to reach the Chandrasekhar mass
and drive the collapse of a central black hole, \cite{Kouvaris:2011fi,McDermott:2011jp}.
Stellar physics can thus provide a strategy to indirectly
test this class of models.
%MT removed: the ADM paradigma. 
Interestingly, the evolution of Main Sequence,
solar mass stars
placed in environments with higher ADM densities 
has not been studied yet (while this 
paper was being refereed, a recent study
addressing very low-mass stars in ADM has 
been published, \cite{Zentner:2011wx}).
Here we address this topic, showing
that the energy transport induced by
ADM cumulating inside such stars can provoke 
dramatic effects on the stellar structure.
%In particular we describe the formation of a small electron degenerate core inside the star already during the Main Sequence.

Whether DM particles are actually intrinsically ``asymmetric'',
or if their self-annihilation rate in stellar environment is low enough
($\sigv\lesssim$10$^{-33}$cm$^3$/s for the Sun) is indistinguishable
for its effects in stellar evolution, \cite{Taoso:2010tg};
hereafter our definition of ``asymmetric'' 
embraces the previous condition.

%$$$$$$$$
{\it ADM in stars---}
%$$$$$$$$$
Here we adopt the same formalism as in \cite{Taoso:2010tg},
to which we address the reader for
details. A thorough description of
the underlying theory can be
found  also in \cite{Scott:2008ns} 
and references therein.
DM particles
can be captured by a star via scattering off nuclei,
the evolution of the total number of DM particles $N_\chi$
inside the star reading:
\begin{equation}
\dot{N_{\chi}}=C-2A N_{\chi}^2- E N_{\chi} 
\label{diffequation}
\end{equation}
where $C$ is the particle capture rate over the star, 
$A$ is the annihilation rate and
$E$ the evaporation one.
For the asymmetric DM candidates we are considering, the annihilation
rate is null and evaporation is negligible 
for the physical conditions on which we focus,
equation \ref{diffequation} therefore 
becomes trivial for ADM.
Here we recall the linear dependence of the capture rate on the
spin-dependent scattering cross section $\sigmsd$ 
and the environmental DM density $\rhochi$.
Captured particles keep on scattering with
the stellar gas reaching a thermally
relaxed distribution inside the star,
$n_{\chi}(r)=n_{\chi,0} e^{-r^2/r_{\chi}^2}.$
For the
DM masses and cross sections we are
going to consider, in a Sun-like star the thermal DM radius $\rchi$
is of the order of $\rchi\lesssim$10$^{9}$ cm ($\sim 0.03$ $R_{\odot}$),
thus making DM particles confined within the 
nuclear energy generation region.
DM particles weakly interacting with the baryons provide
an energy transport mechanism, additional to
the standard ones. 
In their orbits, DM particles  scatter 
in the innermost regions 
(called the ``inversion core'' hereafter)
absorbing the heat and releasing it in the immediate surroudings,
within the nuclear energy production region.

%%%%%%%%%%%%%%%%%%%%%%%%%% FIGURE 1 %%%%%%%%%%%%%%%%%%%%%%%%%%%%%%%%%
\begin{figure}
\centering \epsfig{figure=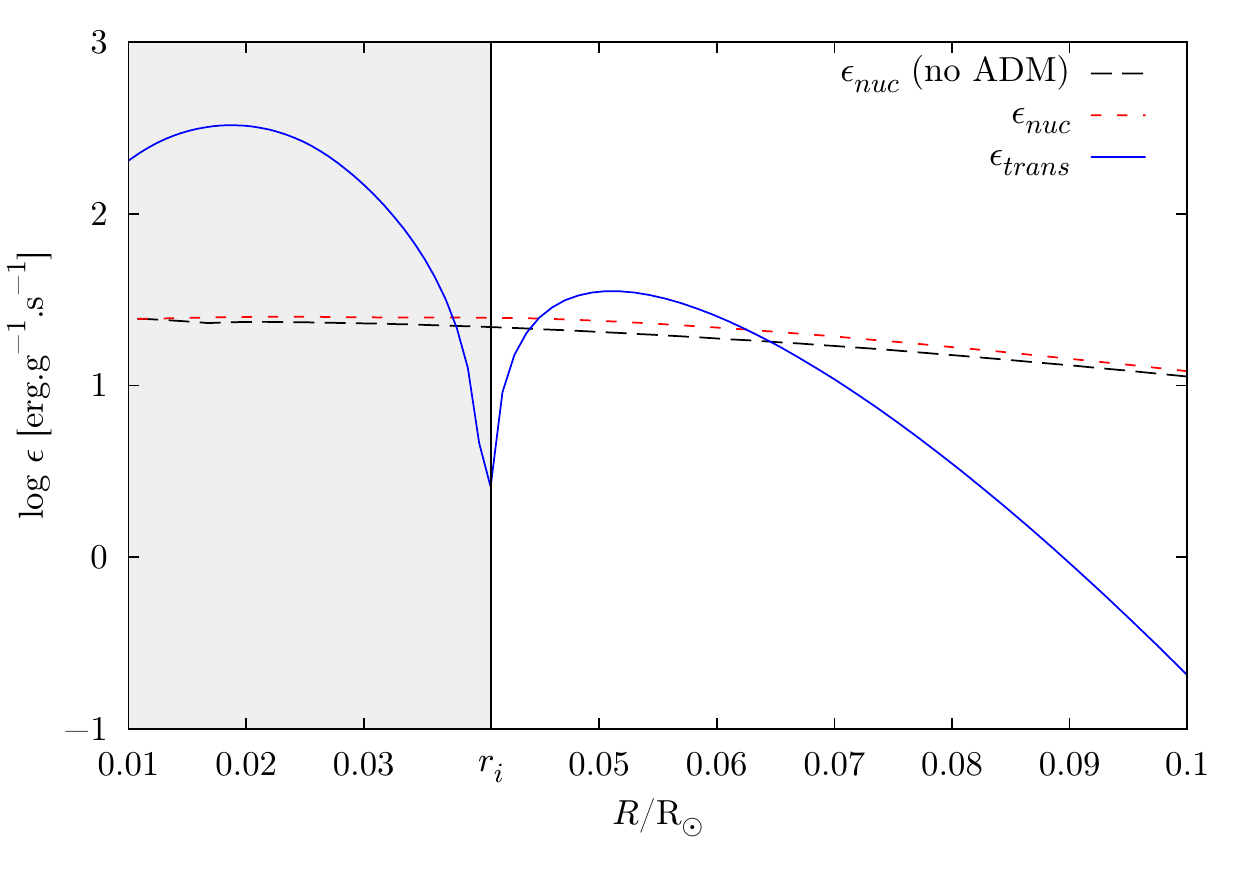,width=1.0\columnwidth,angle=0}
\caption{The structure of $\epstrans$ and $\epsnuc$ for 1M$_\odot$
star evolved with the ADM parameters 
$\rhochi$=10$^3$GeV/cm$^3$, $\sigmsd$=10$^{-37}$cm$^2$, $\mchi$=10GeV.
The profile is taken when the central hydrogen fraction, {\rm X$_c$}=0.2.
Also shown for comparison is
$\epsnuc$ of a star evolved without ADM, at the same age.
The shaded area -innermost of $r_i$- shows where
$\epstrans$ has negative sign, thus absorbing
energy from the stellar gas.
%GM4 note it would be more interesting to compare models with same X_c
}
\label{EnComp}
\end{figure}
%%%%%%%%%%%%%%%%%%%%%%%%%%%%%%%%%%%%%%%%%%%%%%%%%%%%%%%%%%%%%%%%%%%%%%
%$$$$$$$$
{\it Stars in high ADM densities---} 
%$$$$$$$$
The effects of the DM energy transport 
is enhanced for increasing DM densities since
the number of particles cumulated inside the star
grows linearly with $\rhochi$ (see the Appendix for a short discussion
on this).
Hence, we have studied the evolution of
a solar type star (1M$_\odot$, X=0.72, Y=0.266, Z=0.014)
for increasing DM densities. 
In this section we fix the DM-proton spin dependent cross
section $\sigmsd$=10$^{-37}$cm$^2$ and DM 
particle mass $m_\chi$=10GeV.  
We will discuss the effect of varying these parameters in the
following sections, whereas we keep the stellar
velocity through the DM halo $\bar v$=220 km/s.

We have modified the publicly available DarkStars code  \cite{Scott:2009nd},
in order to use the  Spergel \& Press formalism (Eq. 4.9 in \cite{Spergel:1984re})
rather than the Gould \& Raffelt  one, \cite{Gould:1989ez}. 
Whereas the latter has been proved to correct the overestimate
-of order unity-  present in the previous formalism, we have noticed that in critical
conditions such as the ones we meet in our study the
implementation of this formalism may easily induce a
unphysical behavior of the solution obtained by  numerical codes.
We have found this numerical artifact to show up both with the
original DarkStars code and the GENEVA stellar evolution code \cite{Eggenberger:2008eg},
modified for the inclusion of DM effects, as we have described in \cite{Taoso:2010tg}.
The effects we describe in the following could therefore be overestimated of a factor unity,
namely they could show up for DM densities $\rhochi$ a factor unity bigger 
than the actually quoted ones.

We evolve the star from the Zero Age Main Sequence, by
adopting an environmental $\rhochi$=10$^3$GeV/cm$^3$,
which for a NFW DM density profile corresponds to R$_{gal}\sim$10pc
from the Galactic center.
As long as the absolute value of the DM transport energy term $\epstrans$
is much smaller than the nuclear energy generation term $\epsnuc,$ 
the global properties of the star are not modified.
We have verified that  the central temperature 
and density profile {\it are} altered with
respect to the same star evolved without DM, according
to what we have seen in \cite{Taoso:2010tg}:
neutrino fluxes and seismic g-modes
of the star {\it get} modified with respect to
the standard case and we defer a systematic
analysis of this to later studies. 

Here we focus on a much more dramatic effect,
taking place when $\epstrans$ becomes comparable with
$\epsnuc$. For our choice of DM parameters $\mchi$ and $\sigmsd$
this takes place when the total population of DM particles
in the star, $\ntot\sim$5$\times$10$^{47}\#$ (for this
particular choice of parameters and stellar mass,
this takes place approximately 5 Gyr after the Zero Age Main Sequence).
%This happens at an age of approximately 1.7$\times$10$^9$yr for this choice
%of the parameters, as we will see in the following that happens much earlier for
%stars in denser environments
In this conditions DM is able to transport out of a 
very central region
the entire energy generated by nuclear reactions,
thus resulting in an efficient sink of energy for the stellar region within
the inversion radius $r_i\sim$0.04$R_\odot$,
namely the region where the isothermal WIMP temperature is 
lower than the local baryonic temperature,
$\Tchi < \Tb$.
Outside of $r_i$, yet within the nuclear
energy generation region $\rnuc\sim$0.1R$_\odot$, DM particles deposit the 
energy absorbed from the innermost, hotter baryons into the local medium,
colder than $\Tchi$.
This energy deposit is not crucial:
what matters most is the effective diminished efficiency of 
the nuclear energy source at the center.
The star is forced to compensate for this decrease
by increasing the nuclear energy production outside the
inversion core.
%GM9 I would add here
% In Fig. 1 we do not see that but I am wondering whether
% this is due to the fact that the age is the same and
% not Xc. 
%%%%%%%%%%%%%%%%%%%%%%%%%% FIGURE 1 %%%%%%%%%%%%%%%%%%%%%%%%%%%%%%%%%
\begin{figure}
\centering \epsfig{figure=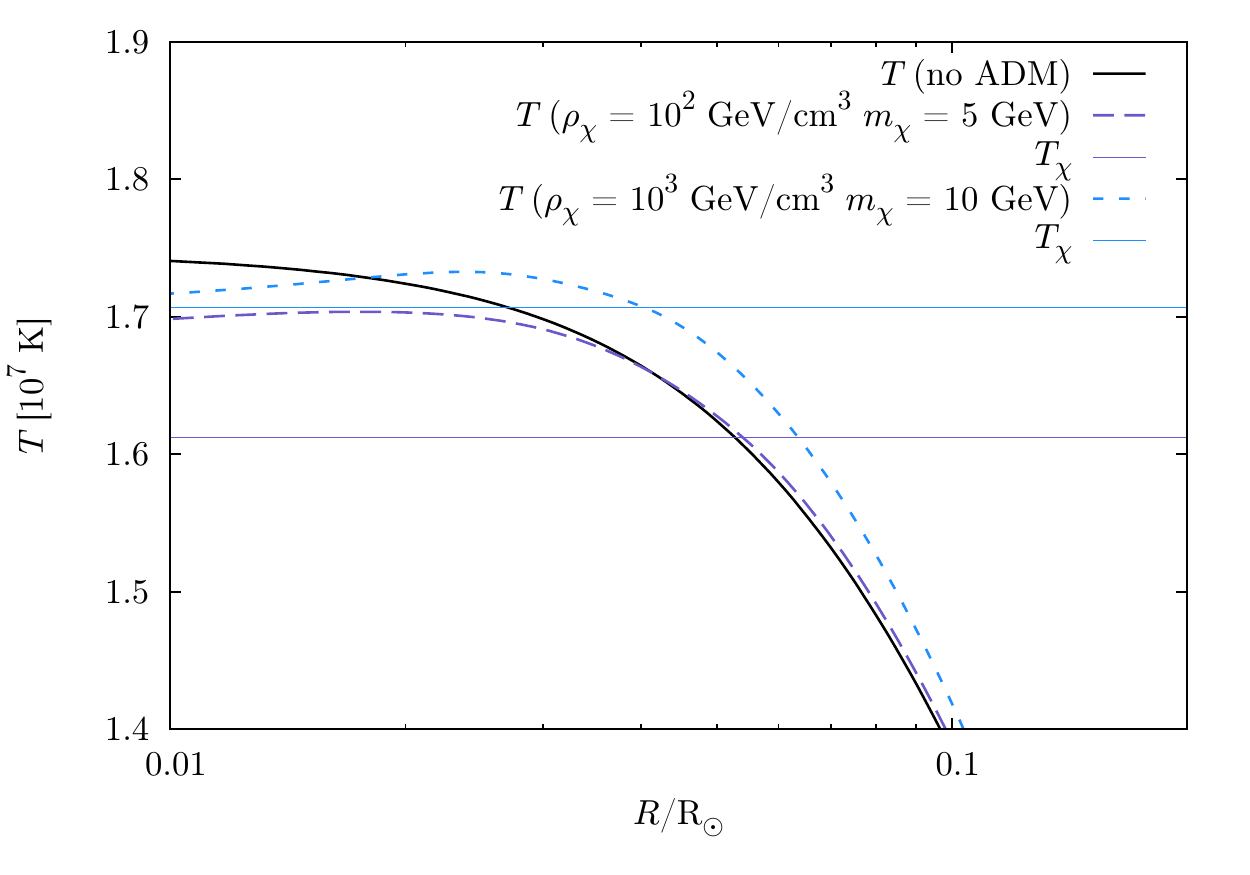,width=1.0\columnwidth,angle=0}
\caption{Stellar (baryonic) temperature profile for two ADM parameter set, as
from the plot; the solid black line is a model evolved without DM.
The profile is taken when the central hydrogen fraction, {\rm X$_c$}=0.2.
This corresponds at an age of approximately 5.3, 5.4, and 5.6 Gyr for 
the case with no ADM, and increasing ADM density $\rhochi$, respectively.
}
\label{StelProf}
\end{figure}
%%%%%%%%%%%%%%%%%%%%%%%%%%%%%%%%%%%%%%%%%%%%%%%%%%%%%%%%%%%%%%%%%%%%%%
In Figure \ref{EnComp} we show the radial structure of
$\epstrans$ compared to $\epsnuc$: it is clearly visible how
the DM energy transport is the mechanism dictating the energetics
of the region within the inversion radius $r_i$, and
the energy sink caused by DM in this region induces
a temperature drop. 
%modulated in turn by the shape
%of $\epstrans$, sensitive to the particle density inside the star,
%which peakes within the thermal WIMP radius
%$\rchi\sim$0.03$R_\odot$.
An example of the resulting temperature profile is shown in
Figure \ref{StelProf}, where both the central drop within
$r_i$ caused by the ADM energy absorption,
and a raise in the shell between $r_i$ and $\rnuc$ are clearly visible.
Such new  temperature profile is the result of the new
equilibrium reached by the star, which readjusts its
structure in order to provide
the correct energetics; the new nuclear
reaction rate $\epsnuc$ is 
visible in Figure \ref{EnComp}.
These modifications are quite remarkable,
modifying the chemical evolution of the star as well as its external
appearance, see more in the next Section.
%-at 6 Gyr from our starting point, the radius of the star is in fact
%$R_*$=1.24$R_\odot$, namely approximately 
%1.06 times the radius of a 1$M_\odot$star
%of the same age, evolved normally without DM-

The behavior we have described is reached when 
DM particles are enough that $\epstrans > \epsnuc$
in the central region of the star, within $r_i$. 
For the value of {\it environmental} DM density
$\rhochi$=10$^3$GeV/cm$^3$ and $\bar v$=220 km/s
adopted in this run, this happens quite
early during the evolution of the star,
whereas for DM densities smaller than 
$\rhochi$=10$^3$GeV/cm$^3$ (with $\mchi$=10GeV),
the time needed for the star to capture enough particles is
longer than the MS itself, so the mechanism we have described
can not take place. 
%as not enough hydrogen will be available
%in the core of the star for Spin-Dependent scatter of DM particles
%when the correct DM densities {\it inside} the star are reached.
Conversely, for increasing DM densities $\rhochi$ (and therefore the
capture rate $C$) this arrives at shorter times during the Main Sequence.

\begin{figure}
\centering \epsfig{figure=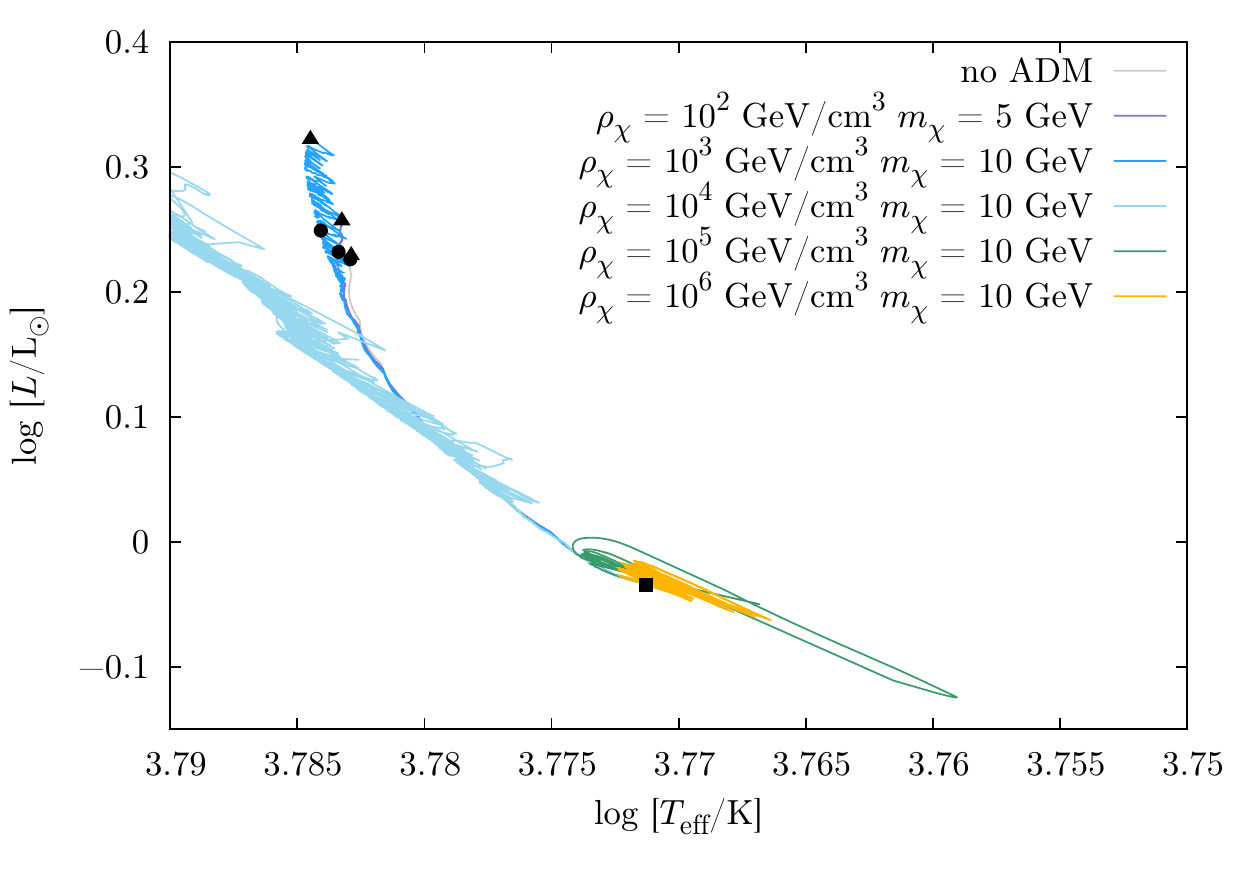,width=1.0\columnwidth,angle=0}
\caption{The Hertzsprung-Russel diagram for 1M$_\odot$, 
for varying DM parameters as marked in the plot, with $\sigmsd$=10$^{-37}$cm$^2$. 
The solid black square is the starting
point of our simulation, black solid circles mark
an age of 6Gyr, not achieved for runs with $\rhochi >$
10$^3$GeV/cm$^3$.
%which stops at an age of 0.33Gyr, with an almost unmodified {\it X$_c$}.
%0.726 vs 0.728 at the beginning
Black solid triangles mark the fall of central
hydrogen fraction {\rm X$_c$} below 0.1. This takes place at an age
of 6.05, 6.36, and 6.39 Gyr for the no ADM case, 
$\rhochi$=10$^2$,10$^3$ GeV/cm$^3$,
respectively. 
Notice that the fluctuations may be numerical artifacts, see Appendix for details.}
\label{HRdiagra}
\end{figure}

We have evolved several stellar models
for different values of the environmental DM density
(keeping constant $\sigmsd$ and $\mchi$), finding that
remarkable effects on the stellar structure are present for 
different DM densities  $\rhochi$. It is worthwhile to remark that
for increasing $\rhochi$, DM particles
start being an effective sink of energy at decreasing times.
The cumulation of more particles during the MS boosts
the $\epstrans$ increasingly, with even more remarkable effects:
in Figure \ref{HRdiagra} we show the impact of
increasing ADM densities on the star's evolution.
%, in consequence of which the 
%$\epsnuc$ falls considerably, and the star
%readjusts contracting.
Whereas for $\rhochi \lesssim$10$^4$GeV/cm$^3$
$| \epstrans | >\epsnuc$ for R$\lesssim \rnuc$,
its integral over the whole star is smaller
than the total nuclear luminosity of the star,
 $\Ltrans\equiv \int | \epstrans | dV < \Lnuc$: the structure reacts to it
by finding a new stable configuration at increased luminosities,
by readjusting the nuclear reaction rate between
$r_i$ and $\rnuc$.
%GMGM I would remove the text below
%By increasing $\rhochi$, $\Ltrans > \Lnuc$ and
%at this point not even the readjustement of the core can
%counteract the effects of DM, as the central
%temperature is kept at an extremely low plateau (but not
%lower than $\Tchi$) by ADM diffusion, and the luminosity
%of the star decreases, as we discuss in the following Section.
%GMGM and replace by the following 
On the other hand, at very high ADM density 
($\rhochi\gtrsim$10$^5$ GeV/cm$^3$)
the temperature is mitigated in the whole 
nuclear energy generation region,
rather than only in the inversion core.
This is because the energy reinjected by the WIMPs at R $> r_i$
is now  sufficient to inflate that region, thus cooling it 
and reducing the efficiency of $\epsnuc$.
The envelope then contracts to extract 
energy from the gravitational potential well
%with the overal final effect to reduce the effective temperature of the star
.

Summarizing, when $\Ltrans < \Lnuc$:
the star compensate for the deficit of nuclear energy
generation in the center by increasing the nuclear energy generation
outside the inversion core. 
At $\Ltrans > \Lnuc$:
the redistribution of the energy by the WIMPS is such that
$\epsnuc$ is reduced in the whole nuclear core and
the star must compensate this deficit by the contraction of
its envelope.

{\it Diagnostic power---}
%$$$$$$$$
Is it possible to use such mechanism as 
a possible probe for ADM parameter space,
namely: are stellar observables affected from this 
mechanism at a testable level?
Deviations from the standard path in the
Hertzsprung-Russel (HR) diagram are the
ideal observable.

The modifications of the stellar structure
described in the previous section, induced by ADM,
are clearly visible in the HR diagram.
In our Figure \ref{HRdiagra} it is possible to see how
the model described above detaches from a normal
path in the temperature-luminosity plane once 
captured DM particles have achieved the right
number $\ntot$. The new distribution of temperature
makes the star more luminous and bigger,
thus moving left and upward in the plane for low ADM densities.
The resulting position of the star
in the HR diagram is therefore unusual,
and more so as the star ages and gets toward the
end of the MS.  
In Figure  \ref{HRdiagra}  it can be appreciated how
by increasing $\rhochi$ the actual position
of the star in the HR changes, finally reverting it to lower
luminosities as a consequence of the dramatic drop
in temperature, as explained in the previous Section;
% and this produces an evolution
% in the HR diagram which is in the opposite direction
% than in the low density regime. Interestingly, when the
% reinjection of the energy by the WIMPS in the surrounding of the
% inversion core is switched off, the star evolve as it evolves
% at lower densities, namely, the star can then compensate
% the lack of nuclear energy production in the centre, by 
% allowing higher temperatures to be reached outside
% the inversion radius. When this energy is injected, on the
% contrary, that region inflates, the temperature decreases
% and the nuclear energy production is reduced in the whole
% nuclear burning core.
still, these stars are kept away from the usual track.
This could indeed be used to 
identify a peculiar generation of
stars, and probe ADM parameters 
or the distribution of ADM in our Galaxy.
However, we caution that these observational tests are challenged by
the difficulty of observing stars close to the galactic center
(or in the very center of Dwarf Galaxies, where
DM environmental conditions could be similar), which
makes current uncertainties on the position of these stars 
in the HR diagram  quite large.
%GM12 note that the positions are strange, provided we know the
% mass and metallicity. These quantities
% should be known for making meaningful comparisons
%with observed position in the HR diagram.

However, it is worth to remark that, unlike
the case of annihilating DM, the effects of
ADM may be ``portable'' by the star throughout
its journey in space. In the case
of DM annihilating and providing an additional
energy source to the star, the effects are
tightly related to the  environmental DM density
the star is experiencing within the short 
equilibrium time -${\mathcal O}$(10$^6$yr). 
Self-annihilation depletes the DM
cumulated inside the star, and a continuous provision is
required to keep the effects going, \cite{Scott:2008ns}. 
ADM does not escape from the star: 
in principle stars may capture ADM in a 
dense environment and migrate somewhere else
in the Galaxy, and the effects would
still be visible.
%%%%%%%%%%%%%%%%%%%%%%%%%% FIGURE 1 %%%%%%%%%%%%%%%%%%%%%%%%%%%%%%%%%
%%%%%%%%%%%%%%%%%%%%%%%%%%%%%%%%%%%%%%%%%%%%%%%%%%%%%%%%%%%%%%%%%%%%%%

%$$$$$$$$
{\it The effect of varying DM parameters---}
%$$$$$$$$
%An extensive and detailed analysis 
%of the parameter space is not the scope of this paper
%and we defer it to further studies.
%Here we discuss the impact of the DM mass $\mchi$ and
%DM-proton Spin-Dependent cross section $\sigmsd$ on
%the results we have obtained.
For the values of $\sigmsd$ we are considering, the DM energy transport
is non local, i.e. the
DM free path is larger than $r_{\chi}$ and DM particles can efficiently
transport energy between distant regions in the stellar core.
In this case, $\epstrans$ is enhanced for larger $\sigmsd$,
until the DM energy transport becomes local and
so $\epstrans$ is decreased, this happening for $\sigmsd\gsim$10$^{-34}$ cm$^2$.
Capture of DM particles is easier for lighter ones,
and heavier DM particles tend to be confined in smaller regions
inside the core, both effects making the DM transport inside stars
more sensitive to small values of $\mchi$.

For $\mchi$=10 GeV and $\sigmsd$=10$^{-38}$cm$^2$ the dramatic
effects on the stellar structure which we have previously discussed
start to appear for $\rhochi$ $\geq$10$^5$GeV/cm$^3.$
This actually demonstrates that smaller SD cross sections can be explored,
although the price to pay is to shrink 
the region of the Galaxy potentially probing this mechanism. 
On the other hand, for $\sigmsd$=10$^{-36}$cm$^2$ 
the same situation is reached at lower DM densities,
$\rhochi$=10$^2$GeV/cm$^3.$
For smaller DM masses
(but still above the evaporation mass $\sim$5 GeV)
it is possible to probe smaller $\sigmsd,$ for
the same value of DM density.
%The different value of $\rchi$ for varying $\mchi$
%also induces differences in the exact redistribution of
%energy (as a conequence of modified $\epstrans$ thus
%resulting in slightly different properties of the star. 

%$$$$$$$$
{\it Effects on massive stars---}
%$$$$$$$$
In principle, the accumulation of ADM particles inside
massive stars should have the same effects as in low-mass ones,
the transport effects of DM not-depending (explicitly) on the burning
mechanism (hydrogen via {\rm p-p} or {\rm CNO}).
However, one crucial fact is to be stressed: 
transport effects start modifying dramatically the structure of a star
when $\epstrans$ is of the same order of $\epsnuc$ 
within $r_i$.
Two important things are to be noticed:
{\it i)} the stellar luminosity scales non linearly with the stellar mass,
$L_* \propto M^{3.5}_*$; 
 {\it ii)} the lifetime becomes 
shorter as the stellar mass increases.
The latter, combined with the fact that the capture rate does {\it not} scale
strongly with the stellar mass (roughly $C\propto M_*$), 
leads to the conclusion that higher mass stars 
are less sensitive than low-mass ones to the same DM parameters.
We have evolved models of increasing stellar
mass between 1 and 10 M$_\odot$ for the same 
ADM parameters, that we take $\sigmsd$=10$^{-37}$cm$^2$, $\rhochi$=10$^7$GeV/cm$^3$,
$\mchi$=5GeV.
The 5M$_\odot$ model
gets out of the typical path on the HR
%$\epstrans$ becomes larger than $\epsnuc$
at an age of approximately 90 Myr ({\rm X$_c\sim$0.2}).
Stars with masses larger than 6M$_\odot$ evolve normally, i.e. the impact
of such ADM densities on their evolution is negligible.

This shows that stars with masses around the solar 
value -or even smaller, as in \cite{Zentner:2011wx}-
are better ADM probes than bigger ones, and yet this
is good news, as approximately 60\% of the stellar mass in our Galaxy 
is expected to be present in the 0.1M$_\odot\leq$M$_*\leq$1M$_\odot$ range.
In the very same environment, low mass stars may be affected by the 
dramatic ADM effects, whereas more massive ones are insensitive to it.
%%Stars in larger regions of the Galaxy could be affected by the
%%mechanism we have described for smaller DM masses 
%%(yet above the
%%evaporation threshold, $\mchi\sim$4GeV for a Sun-like star).

It is also worth discussing the effects produced by this mechanism
on the first stars to form in the Universe, the supposedly
massive Population III. These objects should be naturally placed
in very high DM density environments, as they are born at
the center of a collapsing minihalo at high-redhisft, and this
is known to enhance DM effects onto such stars, \cite{Iocco:2008xb,Taoso:2008kw}.

We have evolved a 100M$_\odot$ star with metallicity Z=10$^{-4}$, 
with increasing ADM densities, finding that for $\sigmsd$=10$^{-37}$cm$^2$
the effects illustrated throughout this paper show up at $\rhochi\geq$5$\times$10$^{10}$GeV/cm$^3$.
Such densities are likely to be achieved in the very central regions of
a primordial star forming halo \cite{Natarajan:2008db};
yet, recent simulations of first star formation show a fragmentation
of the protostellar core, thus yielding to both smaller stellar masses,
and lower central DM densities \cite{Turk:2009ae}.
Whereas no conclusions can be drawn at the level of the state of the art,
it is worth to remark that the two effects may compensate each other,
and PopIII may be indeed affected by ADM.

%$$$$$$
{\it Conclusions---}
%$$$$$$
In this paper we have shown for the
first time that feebly or
non-annihilating DM, carrying weak interactions
with the baryons, can have dramatic effects
on solar mass, Main Sequence stars placed in DM
environments denser than that of our Sun.
This is due to the enhancement of DM-driven
transport effects, that evacuate the
nuclear energy produced in center of 
the core of the star.
We have shown that once this happens 
%(i.e. for DM
%particle masses 5GeV$\lesssim\mchi\lesssim$10GeV and
%spin-dependent cross sections $\sigmsd\gsim$10$^{-37}$cm$^2$)
this may provoke dramatic changes in the structure 
and external appearance of the star.
We have shown that such dramatic effects take place in
solar mass stars in environments with
DM densities $\rhochi\gsim$10$^2$GeV/cm$^3$,
and that they can in principle be used as a probe for DM
particle masses as low as $\mchi$=5GeV and
spin-dependent cross sections $\sigmsd\gsim$10$^{-37}$cm$^2.$
%here Fabio has added sentence on constraints comparison.
Such parameter values are quantitavely comparable with
those that could be placed by the failed observations of
neutron stars in regions with this very same 
(fermionic) ADM density,
see e.g. Figure 1 of \cite{McDermott:2011jp}, yet for as 
difficult as the observation of MS stars can be at the Galactic Center or
ADM dense environments,
they are less challenging than those of cold neutron stars.

%{\bf here something about the modification of chemical signature,
%asteroseismology etc etc, as  we see dramatic modification of the stellar structure.
%Not as dramatic as HR shift, but to be studied in further analysis.}

%\smallskip
{\it Acknowledgements---}
We thank P.~Scott for essential insights on the structure of the DarkStars code.
FI has been supported from the European Community research program FP7/2007/2013 
within the framework of convention \#235878; MT from 
the Spanish MICINNs Consolider–Ingenio 2010 Programme under
grant MULTIDARK CSD2009-00064. 
His work is partly supported by the Spanish grants FPA2008-00319 (MICINN),
PROMETEO/2009/091 (Generalitat Valenciana)
and the European Council (Contract Number UNILHC PITN-GA-2009-237920) and
by the Institute of Particle Physics (IPP) Theory Fellowship and
the Natural Sciences and Engineering Research Council (NSERC) of Canada). 

%%%%%%%%%%%%%%%%%%%%%%%%%%%%%%%%%%%%%%%%%%%%%%%%%%%%%%%%%%%%%%%%%%%%%%
%%%  Bibliography  %%%%%%%%%%%%%%%%%%%%%%%%%%%%%%%%%%%%%%%%%%%%%%%%%%%
%%%%%%%%%%%%%%%%%%%%%%%%%%%%%%%%%%%%%%%%%%%%%%%%%%%%%%%%%%%%%%%%%%%%%%

%$$$$$$
\begin{center}
{\bf Appendix}
\end{center}
%$$$$$$

%$$$$$$
{\it Capture rate---}
%$$$$$$
It is to be noticed that we compute the capture rate 
compute for the value $\bar v$=220 km/s, and the
$\rhochi$ values we quote in the paper follow
this particular choice  for $\bar v$ (typical
of galactic velocity dispersion around the solar radius).
However, the capture rate grows for decreasing $\bar v$,
and it may therefore be useful to think the capture rate in terms
of enhancement with respect to the solar capture; environments 
with $\bar v$ lower than the one we adopt can in fact be easily found,
e.g. above all: Dwarf Galaxies. 
For  $\rhochi$=10$^2$GeV/cm$^3$, $\bar v$=220 km/s,
the enhancement with respect to
the solar capture is  $\sim$250, full formulae can be found in the Appendix
of \cite{Taoso:2010tg}. Self-interaction of DM
particles can also provide a ``boost'' to capture, 
in \cite{Zentner:2009is}.
%$$$$$$
{\it Numerical issues---}
%$$$$$$
In spite of our efforts, we have not been able to
fully understand the reason of the oscillations
seen in Figure \ref{HRdiagra}, whether they are
a physical effect arising from the ``bouncing'' of the
central temperature on the WIMP temperature floor,
or a numerical artifact. We have however checked that
the existence of oscillations does not affect our results
nor does change our conclusions. 
Beside the theoretical consistency of the intepretation presented in the paper, 
one can be convinced of the actual physical consistency of our results
with observations based on several numerical experiments we have performed:
%once one has noticed that the amplitude of oscillations in the HR diagram vary with the
%refinement of the timestep adopted (shorter timestep, smaller oscillations):
{\it a)}, the deviation from the usual HR for mild DM densities takes place {\it before}
oscillations start; {\it b)}, in spite of the variation with timestep of the oscillations
amplitude (shorter timestep, smaller oscillations), the tracks at varying timestep
all converge in the same region of the HR; {\it c)}, no mass loss is associated with oscillation,
therefore excluding the hypothesis that they may drive the evolution of the object
by (artificially, if they are a numerical artifact) changing its mass. 

\end{document}